# A Distributed Architecture for Edge Service Orchestration with Guarantees

Technical Report

Gabriele Castellano[†*]    Flavio Esposito[†]    Fulvio Risso[*]
[†]*Computer Science, Saint Louis University, USA*
[*]*Computer and Control Engineering, Politecnico di Torino, Italy*
Email: [†]{gabriele.castellano, flavio.esposito}@slu.edu, [*]{gabriele.castellano, fulvio.risso}@polito.it

*Abstract*—The Network Function Virtualization paradigm is attracting the interest of service providers, that may greatly benefit from its flexibility and scalability properties. However, the diversity of possible orchestrated services, rises the necessity of adopting specific orchestration strategies for each service request that are unknown a priori. This paper presents *Senate*, a distributed architecture that enables precise orchestration of heterogeneous services over a common edge infrastructure. To assign shared resources to service orchestrators, Senate uses the Distributed Orchestration Resource Assignment (DORA), an approximation algorithm that we designed to guarantee both a bound on convergence time and an optimal (1-1/e)-approximation with respect to the Pareto optimal resource assignment. We evaluate advantages of service orchestration with Senate and performance of DORA through a prototype implementation.

## I. INTRODUCTION

The emerging Network Function Virtualization (NFV) paradigm aims at replacing middleboxes with interconnected virtual functions, easier to deploy, manage and scale [1], [2]. Nowadays, service providers extended this mechanism to orchestrate a large variety of services, thus introducing several challenges, especially at the edge of the network, ranging from finding a feasible placement on the edge infrastructure for each requested service, to managing the services during their life-cycle. These challenges are exacerbated by the diverse (or even federated) orchestration strategies that this variety of services often requires, with different and potentially divergent objectives. Furthermore, different tenants and applications may need personalized policies.

Despite the complexity of this ecosystem, orchestrators have been often designed soundly but as a one-size fits-all (logically) centralized entity [3], [4], handling incoming service requests individually, hence failing to optimize specific goals of individual services. To support recent edge computing architectures and applications, a decentralized orchestration instead would enable smaller-scoped orchestrators to cooperate (or even compete) while deploying multiple services over a shared edge infrastructure. Coordinating such a pool of orchestrators introduces, however, a set of new challenges: how could several orchestrators, each operating with different goals and policies, converge to a globally optimal resource management over a shared edge infrastructure? How could we avoid violations of global policies or feasibility constraints of several coexisting services? How can we guarantee convergence to a distributed agreement and performance optimality given the (often NP-hard) nature of the service placement problem? To answer these questions, we present *Senate*, an architecture for edge service infrastructure orchestration that is decentralized and enables diversification of strategies (i.e., policies). Leveraging the max-consensus literature and the theory of submodular functions, we designed a mechanism in which a set of distributed service orchestrators selectively gossip to reach an agreement on which edge resource has to be (temporary) assigned to which orchestrator. In particular, our contributions are as follows:

**Architectural contributions.** We introduce the *Orchestrators-Resources Assignment problem*, and use linear programming to model its objective and constraints; then, we use the model to define the Senate architecture and define a set of mechanisms that solve the problem with a fully distributed approach (Sections III and IV).

**Algorithmic contributions.** To enable agreement on distributed orchestration leading to an optimal system state, we propose a fully *Distributed Orchestration Resource Assignment* (DORA) algorithm, with guarantees on both convergence time and expected resource assignment performance, even in presence of (non-byzantine) failures (Sections V, VI and VII).

**Evaluation contributions.** We evaluated both *(i)* advantages introduced by Senate service dedicated-orchestration over the traditional one-size fits-all approach and *(ii)* convergence and performance of DORA comparing a few representative system policies (Section VIII).

Finally, we conclude in Section IX.

## II. RELATED WORK

**Orchestration of a large variety of services.** Widespread orchestrators [1], [2], focuses on the architectural solution, without providing details on the objective to be optimized during the service embedding, since this may depend on the service itself. Some existing solutions allow the selection of the orchestration approach that fits best with the service needs [3], [4]. To this end, [5] analyzes the common embedding objective used by existing orchestration solution (e.g., minimization of power consumption, number of physical nodes used, etc.). However, in general it is not possible to reduce the needs of a specific service to one of these generic policies, since *(i)* each service may want to optimize completely different metrics and *(ii)* the desired objective function may often depend on service specific parameters (e.g., the object miss rate in a CDN service). To the best of our knowledge, this is the first work that proposes an approach where the orchestration operates at service granularity.

**Distributed resource assignment.** Mesos [6] is a datacenter-oriented solutions that enables distributed decisions on resource sharing, where a master component makes offers to demanding services. However, mandating the existence of such a component may not be suitable in a scenario where services are executed at the edge of the network. In the above scenario, we should rely on solutions that provide decentralized consensus (e.g., Paxos [7] and Raft [8]) to reach agreement on resource assignment. However, none of them simultaneously provides *(i)* guarantees on convergence time and performance, and *(ii)* a fully distributed approach.

## III. PROBLEM DEFINITION AND MODELING

In this section we define the (NP-hard) *orchestrator-resources assignment problem* by leveraging linear programming. We then use our model in Section IV to design our edge orchestration architecture; in Sections V and VI we present DORA, a distributed approximation algorithm as solution to the centralized problem defined below.

Let us assume that $N_o$ *orchestrators*, operating on a shared edge infrastructure, wish to deploy a bundle (a subset) of $N_s$ services. A service is a virtual instance of an edge function, e.g., a load balancer, a firewall, a traffic accelerator or a point of presence for caching. Each service can be of course implemented by instantiating different edge functions. We assumed that the underlying hosting infrastructure supports $N_f$ distinct edge functions. Furthermore, the infrastructure is partitioned in $N_v$ hosting edge nodes with potentially different physical capacities. Each edge function consumes a given amount of heterogeneous resources that we model with $N_\rho$ different types. Resources are CPU, storage, memory, outgoing bandwidth on a network interface, etc.

Each orchestrator attempts to allocate on the infrastructure all edge functions that are needed by its service bundle. Our goal is to maximize a global utility $u$ while finding an infrastructure-bounded orchestrator-resources assignment that allows the deployment of each service bundle. We define an orchestrator-resources assignment to be *bounded* if the cost of all assigned functions allocated on a hosting node does not exceed the $\boldsymbol{\rho}_n$ available resources on that node.

We model the (centralized) orchestrator-resources assignment problem with the following integer program; each binary decision variable $x_{ijn}$ indicates whether to orchestrator $i$ has been assigned an instance of the edge function $j$ on edge node $n$ or not:

$$\text{maximize} \quad \sum_{i=1}^{N_o}\sum_{j=1}^{N_f}\sum_{n=1}^{N_v} u_{ijn}(\boldsymbol{x}_i)x_{ijn} \quad (1.1)$$

subject to

$$\sum_{i=1}^{N_o}\sum_{j=1}^{N_f} x_{ijn}c_{jk} \leq \rho_{nk} \qquad \forall k \in \mathcal{K},\ \forall n \in \mathcal{N} \quad (1.2)$$

$$\sum_{j=1}^{N_f}\sum_{n=1}^{N_v} x_{ijn} = \sum_{m=1}^{N_s}(\sigma_{im})y_i \qquad \forall i \in \mathcal{I} \quad (1.3)$$

$$\sum_{j=1}^{N_f}\left(\sum_{n=1}^{N_v} x_{ijn}\right)\lambda_{mj} \geq y_i \qquad \forall m \in \mathcal{M},\ \forall i \in \mathcal{I} \quad (1.4)$$

$$\sum_{n=1}^{N_v} x_{ijn} \leq 1 \qquad \forall j \in \mathcal{J},\ \forall i \in \mathcal{I} \quad (1.5)$$

$$\sum_{j=1}^{N_f} x_{ij} \geq 1 - N_f y_i \qquad \forall i \in \mathcal{I} \quad (1.6a)$$

$$\sum_{j=1}^{N_f} x_{ij} \leq 1 N_f y_i \qquad \forall i \in \mathcal{I} \quad (1.6b)$$

$$x_{ijn} \in \{0,1\} \qquad \forall (i,j,n) \in \mathcal{I} \times \mathcal{J} \times \mathcal{N} \quad (1.7a)$$
$$y_i \in \{0,1\} \qquad \forall i \in \mathcal{I} \quad (1.7b)$$
$$c_{jk} \in \mathbb{N} \qquad \forall (j,k) \in \mathcal{J} \times \mathcal{K} \quad (1.7c)$$
$$\rho_{nk} \in \mathbb{N} \qquad \forall (n,k) \in \mathcal{N} \times \mathcal{K} \quad (1.7d)$$
$$\lambda_{mj} \in \{0,1\} \qquad \forall (m,j) \in \mathcal{M} \times \mathcal{J} \quad (1.7e)$$
$$\sigma_{im} \in \{0,1\} \qquad \forall (i,m) \in \mathcal{I} \times \mathcal{M} \quad (1.7f)$$

where $\boldsymbol{x}_i \in \{0,1\}^{N_f \times N_v}$ is the *assignment vector* for orchestrator $i$, whose $j^{th} \times n^{th}$ element is $x_{ijn}$. The auxiliary variables $y_i$ are equal to 1 if at least an instance of any edge function has been assigned to orchestrator $i$, and 0 otherwise (constraints 1.6a, 1.6b, 1.7b). The index sets are defined as $\mathcal{I} \triangleq \{1,\ldots,N_o\}$, $\mathcal{M} \triangleq \{1,\ldots,N_s\}$, $\mathcal{J} \triangleq \{1,\ldots,N_f\}$, $\mathcal{K} \triangleq \{1,\ldots,N_\rho\}$ and $\mathcal{N} \triangleq \{1,\ldots,N_v\}$. The variable $\rho_{nk}$ represents the total amount of the resource $k$ available in the hosting node $n$, hence we name $\boldsymbol{\rho}_n \in \mathbb{N}_n^N$ the resource vector of node $n \in \mathcal{N}$. With $c_{jk} \in \mathbb{N}$ instead we capture the cost of function $j$ in terms of amount of $k$-resource, hence we name $\boldsymbol{c}_j \in \mathbb{N}_\rho^N$ the cost vector of function $j \in \mathcal{J}$. We set $\lambda_{mj} = 1$ if the virtual service $m$ can be implemented through the edge function $j$, while $\sigma_{im} = 1$ if the virtual service $m$ is part of the service bundle of orchestrator $i$.

The *utility function* models the gain $u_{ijn}(\boldsymbol{x}_i)$ obtained by assigning to orchestrator $i$ the edge function $j$ on the edge node $n$, being $\boldsymbol{x}_i$ its current assignment vector. Note that the gain does not depend merely by the service itself, but an orchestrator may benefit differently depending on *(i)* which function is used to instantiate a specific service and *(ii)* on which edge node the chosen function is deployed. Note how constraint (1.2) ensures that the solution is infrastructure-bounded, while constraints (1.3, 1.4) avoid allocation of partial bundles: if orchestrator $i$ get at least a function, then for (1.3) it takes a number of functions equal to its service bundle size, and for (1.4) it also takes, for each service $m$ of its bundle, at least a function able to implement it. Note that equation (1.4) is not enough without (1.3), since if an orchestrator takes an edge function $j$ that is able to instantiate two distinct services $m'$ and $m''$ out of its bundle (e.g., in case a bundle requires two different instances of the same virtual function), constraint 1.4 would be satisfied despite the fact that the orchestrator is missing the additional function to implement either $m'$ or $m''$.

Constraint (1.5) prevents an orchestrator to get multiple instances of the same function on different nodes, situation that should be avoided since it is not possible even on the same node due to the fact that $x_{ijn}$ is not integer, but boolean. Note that this assumption does not constitute a limitation of the model, since the problem instance may contain multiple indexes $j$ for the same physical function, thus allowing an orchestrator to pick it more than once.

## IV. Edge Orchestration with Senate

This section introduces our Senate system architecture (Figure 1). Senate implements the problem defined in Section III in a fully distributed approach, where each *service orchestrator* is aware just of the portion of solution (i.e., edge function embedding) under its responsibility.

As shown in Figure 1, the orchestration system accepts requests modeled as bundles of services, which come with additional information having the purpose to drive the orchestration (i.e., resource allocation, placement and life-cycle management) in a way that is optimal for that specific request (e.g., deployment of a set of content caches for a CDN provider). Metadata that comes with the service bundle are: *(i)* the private utility (i.e., an objective function to be optimized during orchestration), *(ii)* any additional constraint to which the specific service request is subjected to (these constraints complete the model of Section III, since it is not service specific), *(iii)* the preferred embedding heuristic. In the reminder of the paper, we will refer to these information as service orchestration strategies, or *service policies*. Through the *Service-Orchestrator Compiler* component, that takes as input both service policies and *system policies*, an orchestrator for that bundle is generated and run. Then, through the APIs of a (distributed) edge controller, each orchestrator deploys and manages its service bundle over the shared infrastructure. This is partitioned in multiple edge nodes each with different physical capacities.

Details regarding the orchestration compiler, i.e. how service and system policies are composed to generate an orchestration strategy, are out of the scope of this paper. The (automatic) synthesis of the orchestrator is an orthogonal problem that will be presented in a separated work.

As shown in Figure 1, before operating on the shared infrastructure, all generated orchestrators join an overlay through which, using the DORA algorithm, they reach agreement on how resources should be assigned among them. Senate requires peers authentication to access the DORA "agreement layer", thus preventing malicious orchestrators to attack the protocol. The orchestrator's utility (policy) is private and its value may change depending on the edge functions chosen to implement the services, and their placement.

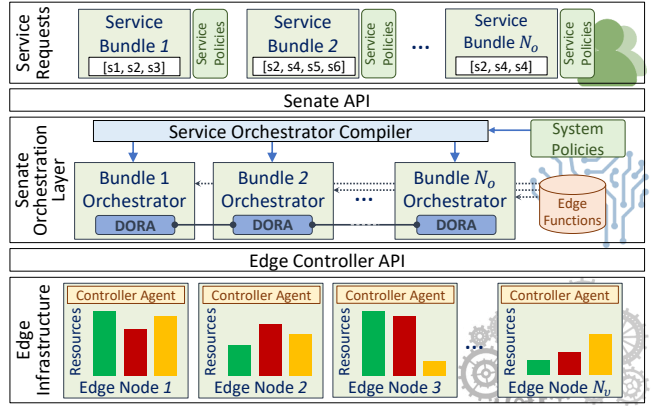

Figure 1. Overall Senate Architecture: DORA processes allow distributed orchestration of multiple edge functions.

Through the employment of two policy layers, Senate enables each orchestrator to optimize its service bundle without leading the overall system to a sub-optimal state. The local embedding heuristic of each orchestrator stretches to optimize the deployment from the service point of view (e.g., selecting the best functions on the best nodes). Then the orchestrator attempts to obtain the resources needed to deploy the selected solution by voting it and taking part to a distributed election, driven instead by the system policies. Agreement on the election results is reached through a consensus-based process (details in Sections V and VI).

In some cases, an orchestrator may need to select a solution that is suboptimal in terms of its service, to obtain the agreement of its peers and therefore the "permission" to deploy its bundle over the infrastructure. This may happen because *(i)* shared resources are running out and other orchestrators introduce better utilities or *(ii)* its private policies diverge from the system ones, so a trade off have to be found.

The next two sections expose in details our DORA algorithm, describing first a single edge-node version, that solves our problem for infrastructures where all shared resources are located on a single hosting node, and then we generalize our solution for a multi-node infrastructure.

## V. Single Edge-node DORA Algorithm

In this section we introduce our Distributed Orchestration-Resource Assignment (DORA) algorithm, starting from a Single Edge-node version (SE-DORA), namely, a single edge node is voted during the election round.

In every version of DORA, each orchestrator $i$ votes for the resources needed for its assignment vector $\boldsymbol{x}_i$, and participates to a resource election protocol. Orchestrators that are elected gain the right to allocate the demanded amount of (virtual) resources on a certain (physical) edge node. In a first phase, each orchestrator performs the election locally, based on known data. Then, a distributed agreement phase leads peers to compute the election results. Note how several existing leader election protocols (e.g. [8], [9]) are based on auctions (they can often be reduced from the generalized auction problem [10]) as they assume that a single item can only be assigned to a single player.

We begin by presenting a simplified version of our DORA algorithm, SE-DORA, to describe all core mechanisms of our approach. Note that, in SE-DORA, structures introduced in Section III are simplified by the absence of the node index $n$.

To describe the SN-DORA algorithm, we need the following definitions:

**Definition 1.** *(private utility function $\boldsymbol{u}_i$).* *Given a set of orchestrators $\mathcal{I}$ and a set $\mathcal{J}$ of $N_f$ functions, we define private utility function of orchestrator $i \in \mathcal{I}$, and we denote it with $\boldsymbol{u}_i \in \mathbb{R}_+^{N_f}$, the utility that orchestrator $i$ gains by adding edge function $j$ to its assignment vector $\boldsymbol{x}_i$; $\mathbb{R}_+^{N_f}$ represents a vector of positive real numbers with size $N_f$.*

Each orchestrator may have a different objective (aside from having no incentive to disclose its utility); however, our assignment algorithm seeks a Pareto optimality where agents seek a global objective maximization despite competing for resources. This global objective is defined as follows.

**Definition 2.** *(global utility function $\boldsymbol{U}$).* *Given a set $\mathcal{I}$ of $N_o$ orchestrators and a set $\mathcal{J}$ of $N_f$ edge functions, we define global utility function, and we denote it with $\boldsymbol{U} \in \mathbb{R}_+^{N_o \times N_f}$, the utility that the system gains by assigning $c_j$ resources to orchestrator $i$ allowing it to add the function $j$ to its assignment vector $\boldsymbol{x}_i$, with $i \in \mathcal{I}$, $j \in \mathcal{J}$, and $c_j \in \mathbb{N}^{N_\rho}$. $\mathbb{R}_+^{N_o \times N_f}$ represents a matrix of positive real number with size $N_o \times N_f$.*

DORA also needs a vote vector that we define as follows.

**Definition 3.** *(vote vector $\boldsymbol{v}^i$).* *Given a distributed voting process among a set $\mathcal{I}$ of $N_o$ orchestrators, we define a vote vector $\boldsymbol{v}^i \in \mathbb{R}_+^{N_o}$, to be the vector of current winning votes known by orchestrator $i \in \mathcal{I}$. Each element $v_\iota^i$ is a positive real number representing the vote $\iota \in \mathcal{I}$ known by orchestrator $i$, if $i$ thinks that $\iota$ is a winner of the election phase. Otherwise, $v_\iota^i$ is 0.*

Since orchestrators decide in a distributed fashion, they will have different views until an agreement on the election winner(s) is reached; we use the apex $i$ to refer to the vote vector as seen by orchestrator $i$ at each point in the agreement process. During the algorithm description, for clarity, we omit the apex $i$ when we refer to the local vector (the same applies also for any of the following data structures).

**Algorithm 1** SN-DORA for orchestrator $i$ at iteration $t$
1: orchestration($\boldsymbol{v}(t-1)$, $\boldsymbol{r}(t-1)$, $\boldsymbol{\rho}$)
2: **if** $\exists \iota \in \mathcal{I}: v_\iota(t) \neq v_\iota(t-1)$ **then**
3:    send($i'$, $t$), $\forall i' \in \bar{\mathcal{I}}_i$
4: receive($i'$, $t$) : $i' \in \bar{\mathcal{I}}$
5: agreement($i'$, $t$)

**Definition 4.** *(demanded resources vector $\boldsymbol{r}^i$).* *Given an voting process among a set $\mathcal{I}$ of $N_o$ orchestrators on $N_\rho$ different types of shared resources, we define as demanded resources vector $\boldsymbol{r}^i \in \mathbb{N}_+^{N_o \times N_\rho}$, the vector of the resources currently requested by each orchestrator; each element $\boldsymbol{r}_\iota^i \in \mathbb{N}^{N_\rho}$ is the amount of resources requested by orchestrator $\iota \in \mathcal{I}$ with its most recent vote $v_\iota^i$ known by $i \in \mathcal{I}$.*

**Definition 5.** *(voting time vector $\boldsymbol{t}^i$).* *Given a set $\mathcal{I}$ of $N_o$ orchestrators participating to a distributed voting process, we define as voting time vector $\boldsymbol{t}^i \in \mathbb{R}_+^{N_o}$, the vector whose each element $t_\iota^i$ represents the time stamp of the last vote $v_\iota^i$ known by $i \in \mathcal{I}$ for orchestrator $\iota \in \mathcal{I}$.*

We also give the following neighborhood definition.

**Definition 6.** *(neighborhood $\bar{\mathcal{I}}_i$).* *Given a set $\mathcal{I}$ of orchestrators, we define neighborhood $\bar{\mathcal{I}}_i \subseteq \mathcal{I} \setminus \{i\}$ of orchestrator $i \in \mathcal{I}$, the subset of orchestrators directly connected to $i$.*

The notion of neighborhood may be generalized with the set of peers reachable within a given latency upper bound.

We are now ready to describe SE-DORA (Algorithm 1).
**Algorithm Overview.** Each service orchestrator $i$ performs an *Orchestration Phase* (Algorithm 2) where an optimal assignment, if any, is built and voted to participate in the resource election, just as in a senate seat assignment. Votes here are updated and topped as in a distributed auction. If any value of the vote vector $\boldsymbol{v}^i$ is changed, $i$ sends its vectors $\boldsymbol{v}^i$, $\boldsymbol{r}^i$ and $\boldsymbol{t}^i$ to its (first-hop) neighbors, then waits for a response coming from any of them. During the *Agreement Phase*, that we also name *ballot*, all vectors $\boldsymbol{v}^{i'}$, $\boldsymbol{r}^{i'}$ and $\boldsymbol{t}^{i'}$ received by neighbor $i'$ are used in combination with the local values (Algorithm 5), to reach an agreement with $i'$. Note that the assignment vector $\boldsymbol{x}_i$ of each orchestrator does not need to be exchanged. Each orchestrator knows how much resources its peers are demanding, but it is unaware of the details regarding which functions they wish to allocate.

The remainder of this section gives more details on the two main phases of the SE-DORA algorithm.

*A. Orchestration Phase*

After the the initialization of local vectors $\boldsymbol{v}(t)$, $\boldsymbol{r}(t)$ and $\boldsymbol{t}(t)$ for the current iteration $t$ (Algorithm 2, line 2), orchestrator uses Algorithm 2, line 8 to elect the current winners according with the known votes updated at the last

**Algorithm 2** orchestration for orchestrator $i$ at iteration $t$

**Input:** $v(t-1)$, $r(t-1)$, $t(t-1)$, $\rho$, $c$
**Output:** $v(t)$, $r(t)$, $t(t)$
1: **if** $t \neq 0$ **then**
2:     $v(t)$, $r(t)$, $t(t) = v(t-1)$, $r(t-1)$, $t(t-1)$
3: **do**
4:     $\bar{v}_i = v_i(t)$
5:     **if** $v_i(t-1) \neq 0 \wedge v_i(t) = 0$ **then**     ▷ outvoted
6:         embedding$(t)$     ▷ find next $x_i$ maximizing $u_i$
7:         voting$(x_i, c)$     ▷ vote $x_i$ using $U$
8:     election$(v(t), r(t), \rho)$
9: **while** $\bar{v}_i \neq v_i(t)$     ▷ repeat until no outvoted

**Algorithm 3** voting for orchestrator $i$ at iteration $t$

**Input:** $x_i$, $c$
**Output:** $v_i(t), r_i(t), t_i(t)$
1: $t_i(t) = t$     ▷ vote time
2: **if** $x_i \neq 0$ **then**     ▷ valid assignment
3:     $r_{ik}(t) = \Sigma_j x_{ij} c_{jk}, \forall k \in \mathcal{K}$     ▷ demanded resources
4:     $v_i(t) = \text{score}(x_i)$     ▷ vote new assignment

**Algorithm 4** election routine at iteration $t$

**Input:** $v(t)$, $r(t)$, $\rho$
**Output:** $v(t)$
1: $\bar{\rho} = \rho$     ▷ residual resources
2: $\mathcal{W} = \varnothing$     ▷ winner set
3: **do**
4:     $\mathcal{I}_b = \{i \in \mathcal{I} |\ r_{ik}(t) \leq \bar{\rho}_k, \forall k \in \mathcal{K}\}$     ▷ candidates
5:     $\omega = \arg\max_{i \in \mathcal{I}_b \setminus \mathcal{W}} \left\{ \frac{v_i(t)}{\|r_i(t)\|} \right\}$     ▷ best voter
6:     $\mathcal{W} = \mathcal{W} \cup \{\omega\}$     ▷ add to winners
7:     $\bar{\rho}_k = \bar{\rho}_k - r_{\omega k}, \forall k \in \mathcal{K}$     ▷ decrease resources
8: **while** $\mathcal{I}_b \setminus \mathcal{W} \neq \varnothing$     ▷ repeat if left candidates
9: $v_\iota = 0, \forall \iota \in \mathcal{I} \setminus \mathcal{W}$     ▷ reset loser votes iteration. After the first iteration, when an orchestrator $i$ has been outvoted (Algorithm 2, line 5), the algorithm starts to alternate between *(i)* an *embedding routine* (Algorithm 2, line 6), which computes the next suitable assignment vector $x_i$ maximizing the private utility, *(ii)* the *voting routine* (Algorithm 2, line 7) where orchestrator $i$ votes for the resources that follow the last computed assignment vector and *(iii)* the *election routine* (Algorithm 2, line 8).

The iteration continues until orchestrator $i$ does not get outvoted anymore (Algorithm 2, line 9). This may happen if either *(i)* the selected assignment vector allows orchestrator $i$ to win the election or *(ii)* there are no more suitable assignments $x_i$ (then no new vote has been generated).

**Remark.** *To guarantee convergence, DORA forbids outvoted orchestrators to re-vote with an higher utility value on resources that they have lost in past rounds. Re-voting is, however, allowed only on residual resources.*

*1) Embedding Routine:* Either during the first iteration ($t = 0$), or any time orchestrator $i$ is outvoted, SN-DORA invokes an embedding routine that, based on the private policies of $i$, computes the next best suitable assignment vector $x_i$. Therefore, this routine is in turn private for each orchestrator, and strictly dependent from the nature of the specific services orchestrated by $i$, thus encapsulating all the ad-hoc "orchestration strategies" discussed in Section IV. In general, we can consider it like an assignment vector builder, that, one by one, adds to $x_i$ the functions that maximize the private utility $u_i(x_i)$, also considering service specific constraints. Therefore, the utility of function added at each step may depend from those that have been already added.

*2) Voting Routine:* After a new assignment vector is build (Algorithm 2, line 7), the orchestrator runs a voting routine, where it updates the time of its most recent vote and then, if the assignment vector is valid, demanded resources are updated and voted, through a *scoring function* that is driven by the global utility (Algorithm 3). Although the raw global utility itself may be used as scoring function to calculate votes, in Section VI-B we give recommendation on which scoring function should be used running DORA to have convergence and optimal approximation guarantees shown in Section VII. Since value $t_i$ is updated in any case (Algorithm 3, line 1), if orchestrator $i$ did not find any other suitable assignment vector, the recent time stamp associated with an empty vote will let its peers know that $i$ agrees with the fact that it definitively lost the election for that node.

*3) Election Routine:* The last step of the *Orchestration Phase* (Algorithm 2, line 8) is a resource election that decides which orchestrators can allocate the demanded resources on the shared edge node (Algorithm 4). Based on the most recent known votes $v(t)$, the related resource demands $r(t)$ and the capacity $\rho$ of the shared edge node, this procedure, adopting a greedy approach, at each step *(i)* discards the orchestrators whose demanded resources $r_i$ exceed the residual capacity and *(ii)* selects the one with the highest ratio vote to demanded resources (Algorithm 4, lines 4-5). The one elected is then added to the winner set and the amount of resources assigned to the new winner are removed from the residual ones (Algorithm 4, lines 6-7). The greedy election ends when either all candidates results winner, or residual resources are not enough for any of those remaining. Finally, votes of orchestrators that did not win the election are reset (Algorithm 4, line 9). As we will see in Section VII, the greedy heuristic gives guarantees on the optimal approximation and has been proved to be the best approximation algorithm for the budgeted max coverage problem, from which we show our problem can be reduced.[1]

### B. Agreement Phase

During this phase, orchestrators make use of a consensus mechanism to converge over their vote vector $v^i$, i.e., reach agreement on which are the winning votes and hence the overall resources assignment (Algorithm 5). By adapting the standard definition of *consensus* [12] to the

---
[1]Authors of [11] show that the standard greedy approach have to be slightly modified (with partial enumeration) in order to preserve guarantees. For brevity, we presented the standard version in Algorithm 4. However, in the average case the standard version computes, faster, the same solutions.

**Algorithm 5** agreement with orchestrator $i'$ at iteration $t$

**Input:** $\boldsymbol{v}(t)$, $\boldsymbol{r}(t)$, $\boldsymbol{t}(t)$, $\boldsymbol{v}^{i'}(t)$, $\boldsymbol{r}^{i'}(t)$, $\boldsymbol{t}^{i'}(t)$
**Output:** $\boldsymbol{v}(t)$, $\boldsymbol{r}(t)$, $\boldsymbol{t}(t)$
1: **for all** $\iota \in \mathcal{I}$ **do**
2:     **if** $t_\iota(t) < t_\iota^{i'}(t)$ **then**     ▷ received newer vote
3:         $v_\iota(t) = v_\iota^{i'}(t)$
4:         $r_{\iota k}(t) = r_{\iota k}^{i'}(t)$, $\forall k \in \mathcal{K}$
5:         $t_\iota(t) = t_\iota^{i'}(t)$

orchestrator-resources assignment problem, we define the *election-consensus* as follows:

**Definition 7.** *(election-consensus). Let's assume a set $\mathcal{I}$ of $N_o$ orchestrators share a computing infrastructure through an election routine driven by, for each orchestrator $i \in \mathcal{I}$, the vote vector $\boldsymbol{v}^i(t) \in \mathbb{R}_+^{N_o}$, the demanded resources vector $\boldsymbol{r}^i(t) \in \mathbb{R}_+^{N_o}$ and the time stamps vector $\boldsymbol{t}^i(t) \in \mathbb{N}^{N_o \times N_\rho}$. Let $e : \mathbb{R}_+^{N_o}, \mathbb{N}^{N_o \times N_\rho} \to 2^\mathcal{I}$ be the election function, that given a vote vector $\boldsymbol{v}$ and the related demanded resources $\boldsymbol{r}$ returns a set of winners. Given the consensus algorithm for orchestrator $i$ at iteration $t+1$, $\forall \iota \in \mathcal{I}$,*

$$v_\iota^i(t+1) = v_\iota^{i'}(t), \; r_\iota^i(t+1) = r_\iota^{i'}(t),$$
$$\text{with } i' = \arg\max_{i' \in \mathcal{I}_i \cup \{i\}} \{t_\iota^{i'}(t)\}, \quad (2)$$

*election-consensus among the orchestrators is said to be achieved if $\exists \bar{t} \in \mathbb{N}$ such that $\forall t \geq \bar{t}$ and $\forall i, i' \in \mathcal{I}$,*

$$\begin{cases} e(\boldsymbol{v}^i(t), \boldsymbol{r}^i(t)) \equiv e(\boldsymbol{v}^{i'}(t), \boldsymbol{r}^{i'}(t)) \\ \boldsymbol{v}_\iota^i(t) \neq 0 \iff \iota \in e(\boldsymbol{v}^i(t)), \; \forall \iota \in \mathcal{I}, \end{cases} \quad (3)$$

*i.e., on all orchestrators the election function computes the same winner set and only winner votes are non zero.*

The agreement for each orchestrator $i$, e.g., on the vector $\boldsymbol{v}^i$, once received $\boldsymbol{v}^{i'}$, $\boldsymbol{r}^{i'}$ and $\boldsymbol{t}^{i'}$ from each $i'$ in the neighborhood, is performed comparing each element $t_\iota$ with $t_\iota^{i'}$ for all $i'$ members of the neighborhood, choosing those with the latest time stamp (Equation 2). Being DORA designed to be an asynchronous algorithm, in our proposal, at each iteration $t$ the agreement can start also if not all vectors from the neighborhood have been received. For each $i'$, if newer information on a given orchestrator vote is received local vectors are updated accordingly (Algorithm 5).

## VI. MULTI EDGE-NODE DORA ALGORITHM

This section extends the SE-DORA algorithm to the multi-node problem. In DORA, each orchestrator may potentially vote simultaneously on different nodes.

We extend our data structure definitions introducing a new index $n \in \mathcal{N}$, where $\mathcal{N}$ is the edge node set and $|\mathcal{N}| = N_v$. For instance, the *vote vector* of Definition 3 becomes $\boldsymbol{v}^i \in \mathbb{R}_+^{N_o \times N_v}$, where each element $v_{\iota n}^i$ is the last vote of orchestrator $\iota \in \mathcal{I}$ on node $n \in \mathcal{N}$ as known by $i$.

Similar to SN-DORA, DORA consists of iteration between an *Orchestration Phase* and an *Agreement Phase*. While *Agreement Phase* in DORA is identical to the single node version, though repeated for each node $n \in \mathcal{N}$, some procedures of the *Orchestration Phase* need to be extended.

During the orchestration phase of DORA, an *embedding routine* selects, for each service of its bundle, both *(i)* the function $j \in \mathcal{J}$ that should be used to implement it and *(ii)* the node $n \in \mathcal{N}$ where $j$ should be placed. This routine may take care of information about nodes that have been lost during the previous elections, according to the private policies (e.g., exclude the lost nodes from the assignment, or use their residual resources for lighter functions).

The *voting routine* is repeated once for each node $n \in \mathcal{N}$ that is involved in the current assignment vector $\boldsymbol{x}_i$, so that a vote $v_{in}(t)$ is generated for each of them. The scoring function, proposed in Section VI-B, allows to generate votes that never exceeds those already elected, thus guaranteeing convergence (details in Section VII).

**Remark.** *Since embedding routine chose functions depending from those already taken, if an orchestrator is outvoted on any node, it resets its votes also on all other nodes, in order build the next assignment vector from scratch.*

According with currently known votes, an election procedure analogous to Algorithm 4 is repeated for each $n \in \mathcal{N}$, so that $N_v$ sets of winners are elected. However, since each orchestrator $i$ needs to be assignee of all voted edge nodes to implement its assignment vector $\boldsymbol{x}_i$, it may happen that one or more winners of a certain node are going to release their vote at the next iteration if they lost at least one needed node. This may possibly lead to suboptimal assignments To cope with this potential suboptimalies, we introduce an *election recount* mechanism, that is run after each election. Details on this subroutine are described in the following subsection.

### A. Election Recount

The election recount mechanism removes from the election *false-winners*, namely, those that won just a subset of the needed nodes, preventing legit peers to win.

Given, for each node $n \in \mathcal{N}$, the set $\mathcal{W}_n \subseteq \mathcal{I}$ of orchestrators that won the election on node $n$, we define the overall winner set $\mathcal{W}^*$ as the union of all the $\mathcal{W}_n$, i.e., $\omega \in \mathcal{W}^*$ if at least a node was won. The election recount subroutine checks, for each $\omega \in \mathcal{W}^*$, if it is not a false-winner, i.e, if it won all needed nodes. This evaluation is performed by recursively checking if $\omega$ lost nodes against other *false-winners*. If exists at least a lost node where the sum of the residual resources and the ones demanded by those false-winners are still not enough for the resources $\boldsymbol{r}_{\omega n}$ demanded by $\omega$, then it is considered a false-winner as well. The winner resolution subroutine returns the set of false-winners, so that the election routine reset all their votes and repeat the elections as long as false-winners are found.

### B. Scoring Function

We now detail the scoring function that each orchestrator uses during the voting routine described in Algorithm 3.

Let $\mathcal{U}_{in}(\boldsymbol{x}_i) = \Sigma_j U_{ijn}(\boldsymbol{x}_i)x_{ijn}$ be the overall node utility of orchestrator $i$ on node $n$. To guarantee convergence of the election process, we let each $i$ communicate its vote on node $n$ obtained from the scoring function:

$$\mathcal{V}_i(\boldsymbol{x}_i, \mathcal{W}_n, n) = \min_{\omega \in \mathcal{W}_n} \{\mathcal{U}_{in}(\boldsymbol{x}_i), \mathcal{S}_{in}(\omega)\}, \quad (4)$$

where $\mathcal{W}_n \subseteq \mathcal{I}$ is the current winner set for node $n$, i.e., $v_{\omega n}(t) \neq 0\ \forall \omega \in \mathcal{W}_n$, and $\mathcal{S}_{in}$ is defined as

$$\mathcal{S}_{in}(\omega) = \begin{cases} +\infty & \text{if } i \text{ never voted on } n, \\ \|\boldsymbol{r}_{in}(t)\|\frac{v_{\omega n}(t)}{\|\boldsymbol{r}_{\omega n}(t)\|} & \text{otherwise.} \end{cases}$$

Since $\mathcal{U}_{in}(\boldsymbol{x}_i) \geq 0$ by definition, if $i$ computes each vote with the function $\mathcal{V}$, it follows that, $\forall (i,n) \in \mathcal{I} \times \mathcal{N}$, $\mathcal{V}_i(\boldsymbol{x}_i, n) \geq 0$. Note how, if it is not the first time that $i$ votes on $n$, the generated vote $v_{in}(t)$ at iteration $t$ never results, during the election process described in Algorithm 4, as an outvote of any previously elected orchestrator for $n$.

Using $\mathcal{V}$ as scoring function leads to the convergence and performance guarantees analyzed in the next section.

## VII. Convergence and Performance Guarantees

In this section we present results on the convergence properties of our DORA algorithm. As defined in Definition 7, by convergence we mean that a valid solution to the orchestrator-resource assignment problem is found in a finite number of steps. Moreover, starting from well known results on sub-modular functions, we show that DORA guarantees an optimal $(1 - e^{-1})$-approximation bound, which also is the best lower bound unless $NP \subseteq DTIME(n^{O(\log \log n)})$.

First, we note that, if (4) is used as scoring function, the election routine of DORA is equivalent to a greedy algorithm attempting to find, for each node $n$, the set of winner orchestrators $\mathcal{W}_n \subseteq \mathcal{I}$ such that the set function $z_n : 2^{\mathcal{I}} \to \mathbb{R}$, defined as

$$z_n(\mathcal{W}_n) = \sum_{\omega \in \mathcal{W}_n} \mathcal{V}_\omega(\boldsymbol{x}_\omega, \mathcal{W}_n, n), \quad (5)$$

is maximized. By construction of $\mathcal{V}$, we trivially have that $z_n$ is monotonically non-decreasing and $z(\varnothing) = 0$.

**Definition 8.** *(sub-modular function). A set function $z: 2^{\mathcal{I}} \to \mathbb{R}$ is sub-modular if and only if, $\forall \iota \notin \mathcal{W}' \subset \mathcal{W}'' \subseteq \mathcal{I}$,*

$$z(\mathcal{W}'' \cup \{\iota\}) - z(\mathcal{W}'') \leq z(\mathcal{W}' \cup \{\iota\}) - z(\mathcal{W}'). \quad (6)$$

This means that the marginal utility obtained by adding $\iota$ to the input set, cannot increase due to the presence of additional elements.

**Lemma VII.1.** *$z_n$ (5) is sub-modular.*

*Proof:* Since $\mathcal{W}'_n \subset \mathcal{W}''_n$, we have

$$\min_{\omega \in \mathcal{W}''_n}\left\{\|\boldsymbol{r}_{\iota n}(t)\|\frac{v_{\omega n}(t)}{\|\boldsymbol{r}_{\omega n}(t)\|}\right\} \leq \min_{\omega \in \mathcal{W}'_n}\left\{\|\boldsymbol{r}_{\iota n}(t)\|\frac{v_{\omega n}(t)}{\|\boldsymbol{r}_{\omega n}(t)\|}\right\},$$

and so, for (4),

$$\mathcal{V}_\iota(\boldsymbol{x}_i, \mathcal{W}''_n, n) \leq \mathcal{V}_\iota(\boldsymbol{x}_i, \mathcal{W}'_n, n). \quad (7)$$

By definition of $z_n$, the marginal gain of adding $\iota$ to $\mathcal{W}_n$ is

$$z_n(\mathcal{W}_n \cup \{\iota\}) - z_n(\mathcal{W}_n) = \mathcal{V}_\iota(\boldsymbol{x}_i, \mathcal{W}_n, n), \forall \iota \notin \mathcal{W}_n \subseteq \mathcal{I},$$

therefore, substituting in (7), we have the definition of sub-modularity (6). ∎

The sub-modularity property of $z_n$ can be intuitively realized noticing that, by definition (4), the scoring function $\mathcal{V}_n$ can, at most, decrease due to the presence of additional elements in $\mathcal{W}_n$.

**Convergence Guarantees.** A necessary condition for convergence on DORA algorithm is that all orchestrators are aware of which are the winning votes for an hosting node. This information needs to traverse all orchestrators in the communication network (at least) once. We show (in Theorem VII.2) that a single information traversal is also sufficient for convergence. The communication network of a set of orchestrators $\mathcal{I}$ is modeled as an undirected graph, with unitary length edges between each couple $i', i'' \in \mathcal{I}$ such that $i'' \in \bar{\mathcal{I}}_{i'}$ and $i' \in \bar{\mathcal{I}}_{i''}$, being $\bar{\mathcal{I}}_{i'} \subseteq \mathcal{I} \setminus \{i'\}$ and $\bar{\mathcal{I}}_{i''} \subseteq \mathcal{I} \setminus \{i''\}$ respectively the neighborhood of $i'$ and $i''$.

**Theorem VII.2.** *(Convergence of synchronous DORA). Given an infrastructure of $N_v$ hosting nodes, whose resources are shared among $N_o$ orchestrators through an election process with synchronized conflict resolution over a communication network with diameter D, if the communications occur over a delay-tolerant channel, then DORA converges in a number of iterations bounded above $N_o \cdot N_v \cdot D$.*

*Proof:* (sketch) Since in DORA each compute node may be assignee to each orchestrator, in the worst case there is a combination of $N_o \cdot N_v$ assignments. From [9] we know that an auction process with synchronized conflict resolution, run by a fleet of $N_u$ processes, where every agent's scoring scheme is sub-modular, agrees on the first $k$ assignments at most by iteration $k \cdot D$. Therefore, in DORA, orchestrators reach agreement on $N_o \cdot N_v$ assignment at most within $N_o \cdot N_v \cdot D$ iterations. Then the claim holds. ∎

As a direct corollary of Theorem VII.2, we compute a bound on the number of messages that orchestrators have to exchange in order to reach an agreement on resource assignments. Because we only need to traverse the communication network at most once for each combination orchestrators per hosting nodes $(i, n) \in \mathcal{I} \times \mathcal{N}$, the following result holds:

**Corollary VII.2.1.** *(DORA Communication Overhead). The number of messages exchanged to reach an agreement on the resource assignment of $N_v$ nodes among $N_o$ non-failing orchestrators with reliable delay-tolerant channels using the DORA algorithm is at most $N_{msp} \cdot N_o \cdot N_v \cdot D$, where D is the diameter of the comunication network and $N_{msp}$ is the number of links in its minimum spanning tree.*

**Performance Guarantees.** Assuming the election routine in DORA being trivially extended with partial enumeration [11], the following results hold:

**Theorem VII.3.** *(DORA Approximation). DORA algorithm extended with partial enumeration yields an $(1 - e^{-1})$-approximation bound with respect to the optimal assignment.*

*Proof:* (sketch) The result in [13] shows that the greedy algorithm modified with partial enumeration is an $(1 - e^{-1})$-approximation for maximizing a non decreasing sub-modular set function subject to a knapsack constraint. DORA assigns resources of node $n$ electing a set $\mathcal{W}_n$ of winning orchestrators through an election routine that, without exceeding the node capacity, adopts a greedy strategy that attempts to maximize the value of the set function $z_n(\mathcal{W}_n)$. Being $z_n(\mathcal{W}_n)$ positive, monotone non-decreasing and sub-modular, hence the claim holds. ∎

**Theorem VII.4.** *(Approximation Bound). The DORA approximation bound of $(1 - e^{-1})$ is optimal, unless $NP \subseteq DTIME(n^{O(\log \log n)})$.*

*Proof:* To prove that the approximation bound given by DORA is optimal, we first show that the orchestrator-resources assignment problem addressed by DORA can be reduced from the (NP-hard) *budgeted maximum coverage problem* [11], defined as follows: given a collection $S$ of sets with associated costs defined over a domain of weighted elements, and a budget $L$, find a subset $S' \subseteq S$ such that the total cost of sets in $S'$ does not exceeds $L$, and the total weight of elements covered by $S'$ is maximized. We reduce the orchestrator-resources assignment problem addressed by DORA from the budgeted maximum coverage problem by considering *(i)* $S$ to be the collection of all the possible set of orchestrators participating to the election process, i.e., $S = 2^{\mathcal{I}}$, *(ii)* $L$ to be the total amount of resources available on the hosting node (in this particular case $N_\rho = 1$), and *(iii)* weight and costs to be votes and demanded resources of each orchestrator. Since in [11] it has been proved that $(1 - e^{-1})$ is the best approximation bound for the budgeted maximum coverage problem unless $NP \subseteq DTIME(n^{O(\log \log n)})$, the claim holds. ∎

## VIII. EVALUATION

To test the proposed distributed orchestration approach, we implemented a prototype of Senate, available at [14]. In our evaluation we assess both the advantages of using service specific orchestrators and the performance of the DORA asynchronous consensus algorithm.

### A. Senate service orchestration

To assess the advantages of Senate service-specific orchestration, we analyze two service requests of a different nature: *(i)* the placement of caches for a CDN provider, and *(ii)* the deployment of a chain of network functions.

**CDN Caches.** A CDN provider provisions caches content over an edge network where user density dynamically changes across edge nodes. The objective of the provider is to minimize the average miss-rate occurring on deployed caches. The orchestrator should react to events where a set of users shifts from a node to another. In our tests we simulated a set of $100$ users moving over a network of $10$ edge nodes managed by a single orchestrator; (the user concentration among nodes has been measured though the Gini index). We summarize our findings in a few take home messages (Figure 2a):

*(i) A generic orchestrator that places caches balancing the resource consumption per node* (Figure 2a, red line), achieves good performance when the concentration index is low (users are well distributed), but the number of miss-rate grows fast when the concentration increases.

*(ii) A generic orchestrator that places caches according with the traffic per node* (Figure 2a, blue line), has poor performance (about $40\%$ miss-rate) when users are well distributed, but achieves optimal miss-rates when users are concentrated on few nodes. This is because a low traffic amount on a certain node does not necessarily mean that users are consuming less number of contents.

*(iii) A service specific orchestrator that decides where to place caches based on the current miss-rate on each nodes* (Figure 2a, green line) achieves a low miss rate both for low and high users per node concentration.

**VNF Chain.** In this set of experiments, a VNF Chain has to be deployed over an edge network between two endpoints, bounding latency below a given threshold. In our tests, we deployed a chain that requires a maximum latency of $50ms$ and contains a firewall whose introduced latency increases with the number of rules that it has to process (Figure 2b). Our findings are summarized as follows:

*(i) A generic orchestrator that minimizes the path latency according with network information* (Figure 2b, red line) has poor worst performance being unaware of the latency introduced by each edge function and its inability to measure improvements introduced when scaling out the chain.

*(ii) A NFV orchestrator that minimizes the sum of path and middle-boxes latency, also scaling VNFs based on the traffic detected on the path* (Figure 2b, blue line) obtains better performance, but it is unable to keep latency under the desired value. This is due to its unawareness of the real latency introduced by the firewall.

*(iii) A service orchestrator that can also scale the firewall based on the current number of rules* (Figure 2b, green line) is able to keep latency under the desired value having a more fine-grain knowledge of its effects.

### B. DORA evaluation

To evaluate DORA performance we prototyped an environment with $4$ edge nodes, each with a different amount of resources (CPU, memory and storage), $9$ available edge functions implementing $6$ diverse services. On average, each function uses about $10\%$ of a node capacity; our tests vary the average service bundle lengths (with averages from $2.4$ to $3.6$). Figures 2c-d show results for three diverse system policies: *(i)* an assignments where the bundle of an orchestrator is entirely allocated in one node is preferred; *(ii)* preference is for assignments in which the orchestrator bundle is distributed on as many node as possible; *(iii)* no preference on the number of nodes is given (private utilities are used to vote).

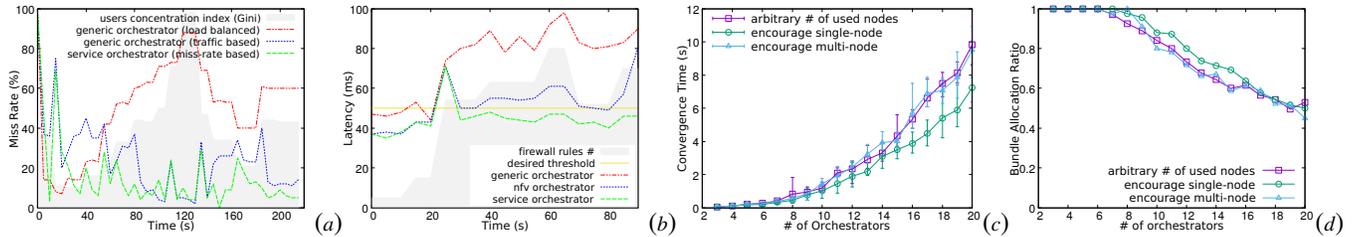

Figure 2. (a) Miss rate over time for a CDN cache provisioning with different orchestrator strategies. (b) Latency over time for a NFV chain provisioning with different orchestration strategies. (cd) Convergence time (c) and bundle allocation ratio (d) of DORA prototype for different system policies.

Figures 2c-d show convergence times and allocation ratios for different policies. For large amount of orchestrators, encouraging single-node solutions leads to significantly lower convergence time (Figure 2c), since a consequences of this policy are *(i)* reduced probability to lose a node ballot and *(ii)* discourage re-voting on residual node resources to place few edge functions (noticerable particularly when resources are running out). As we can see from Figure 2d, this policy also leads to a slightly higher allocation ratio for a certain number of orchestrators. However, this advantage disappears for large amount of orchestrators, since each of them cannot benefit from instantiating small functions on residual resources.

## IX. CONCLUSION

In this paper we proposed *Senate*, a distributed architecture that enables orchestration of heterogeneous services over a shared (edge) infrastructure and provides guarantees on both convergence time and performance. We used linear programming to define and model the orchestrator-resources assignment problem and then presented the core mechanism of *Senate*, a fully Distributed Orchestration Resource Assignment (DORA) algorithm that resolves assignees through a distributed election, and leads to an agreement on the election results. Our evaluation of Senate shows surprising results on the service orchestration approach, and a faster convergence time for a system utility that promote the placement of most part of the bundle on the same node.

## ACKNOWLEDGMENT

The work of Gabriele Castellano was conducted as visiting scholar in the Computer Science Department at Saint Louis University.